\documentclass[aps,prd,
showkeys,showpacs,amssymb,amsfonts,epsf,
preprintnumbers,nofootinbib,superscriptaddress]{revtex4}

\usepackage[dvips]{graphicx}
\usepackage{bm,latexsym,amsmath,amssymb,amsfonts}

\newcommand\beq{\begin{eqnarray}}
\newcommand\eeq{\end{eqnarray}}

\begin{document}

\title{Thick de Sitter brane solutions in higher dimensions}

\author{Vladimir Dzhunushaliev
\footnote{Senior Associate of the Abdus Salam ICTP}}
\affiliation{Arnold-Sommerfeld-Center for Theoretical Physics, Department f\"{u}r Physik, Ludwig-Maximilians-Universit\"{a}t, Theresienstr. 37, D-80333, Munich, Germany\\
and\\
Department of Physics and Microelectronic
Engineering, Kyrgyz-Russian Slavic University, Bishkek, Kievskaya Str.
44, 720021, Kyrgyz Republic}
\email[Email: ]{vdzhunus@krsu.edu.kg}
\author{Vladimir Folomeev}
\affiliation{Institute of Physics of National Academy of Science
Kyrgyz Republic, 265 a, Chui Street, Bishkek, 720071,  Kyrgyz Republi}
\email[Email: ]{vfolomeev@mail.ru}
\author{Masato Minamitsuji}
\email[Email: ]{ minamituzi"at"sogang.ac.kr}
\affiliation{Arnold-Sommerfeld-Center for Theoretical Physics, Department f\"{u}r Physik, Ludwig-Maximilians-Universit\"{a}t, Theresienstr. 37, D-80333, Munich, Germany\\ and\\
Center for Quantum Spacetime, Sogang University,
Shinsu-dong 1, Mapo-gu, 121-742 Seoul, South Korea
}
\begin{abstract}
We present thick de Sitter brane solutions
which are supported by two interacting {\it phantom} scalar fields
in five-, six- and seven-dimensional spacetime.
It is shown that for all cases regular solutions with anti-de Sitter asymptotic (5D problem) and
a flat asymptotic far from the brane (6D and 7D cases) exist.
We also discuss the stability of our solutions.
\end{abstract}

\pacs{04.50.+h, 98.80.Cq}
\keywords{Higher-dimensional gravity, Cosmology}
\preprint{CQUeST-2009-0240}
\maketitle

\section{introduction}

In recent years, there is growing interest in higher-dimensional cosmological models inspired by the recent developments in string theory.
One sort of such models are
branes which are submanifolds embedded into the higher-dimensional spacetime.
They play an important (even essential) role, for instance,
in a description of the confinement of the nongravitational interactions
and/or
stabilization of the extra dimensions.
Braneworld models, in which our universe is exactly
one of four-dimensional branes, are the
central issues in this research field.
Much efforts to reveal cosmology on the brane
have been done in the context of five-dimensional spacetime,
especially after the stimulating proposals by Randall and Sundrum \cite{rs}
(see, e.g., \cite{bw}).
But of course
there is no particular reason to restrict the number of dimensions to be
five.
In recent years, the focus has turned to
higher-dimensional braneworld models.

It is still unclear how the gravity and cosmology on the brane embedded
in the higher-dimensional spacetime behave.
In higher dimensions, there would be many kinds of classes of
possible braneworld models.
Here we focus on the generic feature of higher-dimensional braneworld
models and from this
we deduce the subject to be revealed in this paper.

It is commonly assumed that a brane is an infinitely thin object.
In five dimensions, this is a good approximation
as long as one is interested in the scales larger than the brane thickness.
Then, the cosmology on the brane in five dimensions,
i.e., dynamics of the brane in one bulk spacetime,
is uniquely determined by the Israel junction conditions
once the matter on the brane is specified.
But, in higher dimensions the situation is quite different.
It happens because self-gravity of
an infinitesimally thin three-brane in a higher-dimensional
spacetime develops a severe singularity 
and one cannot put any kind of matter on the brane
if the number of codimensions is larger than 2.
Also in the case of a codimension-two brane, one cannot put
any kind of matter other than the pure tension.
The cosmology on such a brane must be investigated under some prescription
of the singular structure of the brane.
The well-motivated prescription is to regularize the brane by
taking its microscopic structure into consideration.
However, in the case of six-dimensional models,
the cosmology on the brane strongly depends on the way of regularization.
This implies that
there is no unique brane cosmology
in the case of higher codimensions.

Thus, in this paper we take a different point of view.
We shall take the brane thickness into account {\it from the beginning}
and look for the regular braneworld solutions with finite thickness,
called {\it thick brane solutions}, in a field theory model.
The thickness of the brane may be very close to the
length scale of the quantum gravitational theory,
e.g., the string length scale.
As we mentioned, the brane thickness must be more essential in
higher-dimensional spacetimes than in the five-dimensional one.
Thus, we look for the thick brane solutions in higher dimensions
as well as in five dimensions.
Bearing the cosmological applications, we focus on the thick
brane solutions which have de Sitter geometry in the ordinary
four-dimensions.

The properties of thick branes in five-dimensions,
especially localization of gravity, have been discussed
in Ref. \cite{tb_gen}.
In five dimensions,
the thick brane solutions in five-dimensional spacetime
have been explored in various literatures (see, e.g.,
\cite{thick}).
In this paper, we consider thick brane solutions
supported by two interacting scalar fields $\phi$ and $\chi$,
whose potential is given by
\begin{equation}
\label{pot2}
    V(\phi,\chi)=\frac{\lambda_1}{4}(\phi^2-m_1^2)^2+
    \frac{\lambda_2}{4}(\chi^2-m_2^2)^2+\lambda_3 \phi^2 \chi^2+\Lambda.
\end{equation}
A stronger interaction between two scalar fields develops
the local metastable vacua, which may be able to
support our brane universe, other than the global one.
Several works on thick brane solutions in such a setup
have already been performed.
Five-dimensional Minkowski brane solutions were derived in \cite{2scalar1,
phantom}.
Bearing cosmological applications in mind,
our goal in this work
is to find thick de Sitter brane solutions.
Thick brane solutions in higher-dimensional spacetimes
with two interacting scalar field were also investigated
in \cite{2scalar2} (see, e.g., \cite{higher_dim} for other kinds of models).
We shall give de Sitter versions of these solutions.

In Ref.~\cite{gauge} there are some
arguments in favor of the fact
that the scalar fields, used in \cite{2scalar2}, are a quantum nonperturbative condensate of a SU(3) gauge field.
Briefly these arguments consist of the following: components of the SU(3) gauge field can be divided in some natural way onto two parts.
The first group contains those components which belong to a subgroup $SU(2) \in SU(3)$. The remaining components belong
to a factor space $SU(3)/SU(2)$.
Following Heisenberg's idea \cite{heisenberg} about the non-perturbative quantization of a nonlinear spinor field the nonperturbative quantization for the SU(3) gauge field is being carried out as followings.
It is supposed that two-point Green functions can be expressed via the scalar fields. The first field $\phi$ describes two-point Green functions for SU(2) components of a gauge potential, and the second scalar field $\chi$ describes two-point Green functions for $SU(3)/SU(2)$ components of the gauge potential. It is supposed further that four-point Green functions can be obtained as some
bilinear combination of the two-point Green functions. Consequently, the Lagrangian of the SU(3) gauge field takes the form \eqref{Daction}. 
It allows us to regard such sort of thick brane solutions as some defect in a spacetime filled by a condensate of the gauge field living in the bulk.

This paper is organized as follows.
In Sec. II, we introduce the basic field theory model
to find thick brane solutions.
In Sec. III, we present
thick brane solutions in five-, six- and seven-dimensional
models.
In Sec. IV, we investigate perturbations in the five-dimensional
model and then give speculations about stability
of higher-dimensional models.
Secion V contains a brief summary.

\section{Basic theory}

Our interest is in the
$D=(4+n)$-dimensional Einstein-scalar theory whose
action is given by
\begin{equation}
\label{Daction}
S = \int d^Dx\sqrt {- g} \left\{ -\frac{M_{n+4}^{n+2}}{2}R
+\epsilon\left[\frac{1}{2}\partial_A \phi \partial^A
\phi+\frac{1}{2}\partial_A \chi \partial^A
\chi-V(\phi,\chi)\right]
\right\}~,
\end{equation}
where $M_{n+4}$ is the gravitational energy scale
in the spacetime with $n$ extra dimensions.
The potential energy $V$ is defined by Eq (\ref{pot2}).
$\epsilon=+ 1(-1)$ corresponds to the case
of normal (phantom) scalar fields and $\Lambda$ is an arbitrary constant.
As we mentioned in the introduction,
in Ref.~\cite{gauge},
it was argued that
such a theory composed of two interacting scalar fields
coupled to the remaining $U(1)$ degrees of freedom
appears
as a result of
the gauge condensation in the original $SU(3)$ gauge theory.
In this case,
the scalar fields $\phi$ and $\chi$ correspond to the
vacuum expectation values of
the $SU(2)$ and $SU(3)/(SU(2)\times U(1))$ gauge sectors,
respectively.

We assume that
the generalized $D$-dimensional metric
has the static form~\cite{Singleton}
\begin{equation}
\label{metric_n}
ds^2= a ^2(r) \gamma_{\alpha \beta }(x^\nu)dx^\alpha dx^\beta -
\lambda (r) (dr^2 +  r^2 d \Omega ^2 _{n-1})~,
\end{equation}
where $d \Omega ^2 _{n-1}$ is the solid angle for the $(n-1)$ sphere.
$\gamma_{\alpha \beta}$ is the metric of the four-dimensional de Sitter
space whose scalar curvature is given by
$R_{\mu\nu}[\gamma]=3H^2 \gamma_{\mu\nu}$.
The corresponding Einstein equations are given by:
\begin{eqnarray}
\label{Einstein-na}
 &&   3 \left( 2\frac{a ^{\prime \prime}}{a} -
    \frac{a ^{\prime}}{a} \frac{\lambda ^{\prime}}{\lambda }
    \right) + 6 \frac{(a ^{\prime})^2}{a^2}+(n-1)
\left[
3\frac{a ^{\prime}}{a}\left(\frac{\lambda^\prime}{\lambda}+\frac{2}{r}\right)+
    \frac{\lambda^{\prime\prime}}{\lambda}-
    \frac{1}{2}\frac{\lambda ^{\prime}}{\lambda}\left(\frac{\lambda^\prime}{\lambda}-
    \frac{6}{r}\right)+ \frac{n-4}{4}\frac{\lambda^\prime}{\lambda}
    \left(\frac{\lambda^\prime}{\lambda}+
    \frac{4}{r}\right)  \right]
 \\ \nonumber
&-&6\frac{H^2\lambda}{a^2} =-\frac{2\lambda \,\epsilon}{M_{n+4}^{n+2}}
\Big[\frac{1}{2\lambda}\left(\phi^{\prime 2}+\chi^{\prime 2}\right)
+V(\phi, \chi)\Big],
\end{eqnarray}
\begin{eqnarray}
\label{Einstein-nb}
 &&   12 \frac{(a ^{\prime})^2}{a^2 }+
    (n-1) \left[ 4 \frac{a ^{\prime}}{a}\left(\frac{\lambda^\prime}{\lambda}+
    \frac{2}{r}\right)+
    \frac{n-2}{4}\frac{\lambda^\prime}{\lambda} \left(\frac{\lambda^\prime}{\lambda}+
    \frac{4}{r}\right) \right]-12\frac{H^2\lambda}{a^2}
\\ \nonumber
   & =&    -\frac{2\lambda \,\epsilon}{M_{n+4}^{n+2}}
\Big[
-\frac{1}{2\lambda}\left(\phi^{\prime 2}+\chi^{\prime 2}\right)+V(\phi, \chi)
\Big],
\end{eqnarray}
\begin{eqnarray}
\label{Einstein-nc}
 &&   4 \left( 2\frac{a ^{\prime \prime}}{a} -
    \frac{a ^{\prime}}{a} \frac{\lambda ^{\prime}}{\lambda }
    \right) + 12 \frac{(a ^{\prime})^2}{a ^2}+(n-2)
\nonumber \\
   &\times&
\left[ 4\frac{a ^{\prime}}{a}\left(\frac{\lambda^\prime}{\lambda}+
    \frac{2}{r}\right)+
    \frac{\lambda^{\prime\prime}}{\lambda}-
    \frac{1}{2}\frac{\lambda^{\prime}}{\lambda}
    \left(\frac{\lambda^\prime}{\lambda}-\frac{6}{r}\right)+
    \frac{n-5}{4}\frac{\lambda^\prime}{\lambda}
    \left(\frac{\lambda^\prime}{\lambda}+\frac{4}{r}\right) \right]-12\frac{H^2\lambda}{a^2}
\\ \nonumber
&=&  -\frac{2\lambda \,\epsilon}{M_{n+4}^{n+2}}
\Big[\frac{1}{2\lambda}\left(\phi^{\prime 2}+\chi^{\prime 2}\right)+V(\phi, \chi)\Big]~.
\end{eqnarray}
The scalar field equations are
\begin{eqnarray}
  \phi'' + \left(
        \frac{n-1}{r} + 4 \frac{a'}{a} +\frac{n-2}{2} \frac{\lambda'}{\lambda}
    \right) \phi' &=& \lambda \phi \left[
        2 \lambda_3\chi^2 + \lambda_1 \left( \phi^2 - m_1^2 \right)
    \right] ,\\
    \chi'' + \left(
        \frac{n-1}{r} + 4 \frac{a'}{a} +\frac{n-2}{2} \frac{\lambda'}{\lambda}
    \right) \chi' &=& \lambda \chi \left[
        2 \lambda_3\phi^2 + \lambda_2 \left( \chi^2 - m_2^2 \right)
    \right].
\end{eqnarray}
%
%
The extremum of the bulk potential
are located at the following four points
\beq
&\Big(\phi=\pm\phi_0:=\pm
     \sqrt{\frac{2\lambda_3\lambda_2 m_2^2-\lambda_1\lambda_2 m_1^2}
           {4\lambda_3^2-\lambda_1\lambda_2}},
\chi=\pm \chi_0:=\pm
     \sqrt{\frac{2\lambda_3\lambda_1 m_1^2-\lambda_1\lambda_2 m_2^2}
           {4\lambda_3^2-\lambda_1\lambda_2}}
\Big)&
\nonumber\\
&(\phi:=0,\chi:=0)&
\nonumber\\
&
(\phi:=\pm m_1,\chi:=0)
&
\nonumber\\
&
(\phi:=0,\chi:=\pm m_2)
&.
\eeq
We call these points 0, 1, 2 and 3, respectively,
with correspoding potentials $V_i$ ($i=0,1,2,3$).
Clearly, ${\rm max} (V_2,V_3)\leq V_1$.
We also find
\beq
 &&  V_0-V_2
   =\frac{\lambda_1\big(-2\lambda_3 m_1^2+\lambda_2 m_2^2 \big)^2}
         {16\lambda_3^2-4\lambda_1\lambda_2},
\nonumber\\
&&V_0-V_3
   =\frac{\lambda_2\big(-2\lambda_3 m_2^2+\lambda_1 m_1^2 \big)^2}
         {16\lambda_3^2-4\lambda_1\lambda_2},
\nonumber\\
&&V_0-V_1=\frac{\lambda_1 \lambda_2
   \big[m_1^2 (\lambda_1 m_1^2-2 \lambda_3m_2^2)
       +m_2^2 (\lambda_2 m_2^2-2 \lambda_3m_1^2)\big]}
         {16\lambda_3^2-4\lambda_1\lambda_2}.
\eeq
The potential (\ref{pot2}) has two global
minimums 3 and two local minima 2 at
values of the parameters $\lambda_1, \lambda_2$ used in this
paper. The conditions for existence of
local minima are: $\lambda_1>0, m_1^2>\lambda_2 m_2^2/(2\lambda_3)$, and for
global minima: $\lambda_2>0, m_2^2>\lambda_1 m_1^2/(2\lambda_3)$. From
comparison of the values of the potential in global and local minima
we have (taking into account that $V_{loc}=0$ at $V_0=\lambda_2 m_2^4/4$, see below for the six-
and seven-dimensional cases)
$V_{gl}<V_{loc} \Longrightarrow V_{gl}<0$, i.e. $\lambda_2
m_2^4> \lambda_1 m_1^4$. Besides two local and two global minima,
four unstable saddle points (0) and the local maximum (1) exist.
In the next sections, we look for thick brane solutions starting and finishing
in one of local minima 2 of the potential  (\ref{pot2}).

For our purpose, in the following sections
the bulk coordinate and scalar field variables
are rescaled to make them dimensionless as
$r\to H^{-1} r,\, 
\phi\to M_{n+4}^{-(n+2)/2}\phi$ 
and
$\chi\to M_{n+4}^{-(n+2)/2}\chi$.
Correspondingly, the potential and its parameters are also rescaled as
$
V \to H^2 M_{n+4}^{n+2} V$,
$\lambda_i\to \lambda_i H^2/ M_{n+4}^{n+2}$ and
$m_j^2\to M_{n+4}^{n+2} m_j^2$,
where $i=1,2,3$ and $j=1,2$.

Then, the boundary conditions at $r\to 0$ are given by
$a'(0)=0$, $\phi'=0$ and $\chi'=0$.
And from the constraint relation of Einstein equations
Eq. (\ref{Einstein-nb}),
we find the boundary value of scale factor
\beq
a(0)=\sqrt{\frac{6\epsilon }{ V(0)}}. \label{ini}
\eeq
At the asymptotic infinity, the scalar fields approach the local minimum where
$\phi{}'(\infty)=\chi{}'(\infty)=0$.

\section{Thick de Sitter brane solutions}

\subsection{The five-dimensional model}

By substituting $n=1$ and $\lambda(r)=1$ in the metric ansatz Eq.
(\ref{metric_n}), we have
\beq
ds^2=a(r)^2 \gamma_{\mu\nu}dx^{\mu}dx^{\nu}-dy^2.
\eeq
Then, 
the Einstein equations (\ref{Einstein-na}), (\ref{Einstein-nb})
and (\ref{Einstein-nc}) now read
\beq
&&6\Big(\frac{a'}{a}\Big)^2-\frac{6}{ a^2}
=\epsilon
\Big(\frac{1}{2}\phi'{}^2+\frac{1}{2}\chi'{}^2-V\Big),
\nonumber\\
&&
 \frac{a''}{a}
=-\frac{\epsilon}{2}
\Big[
 \frac{1}{2}(\chi')^2
+\frac{1}{2}(\phi')^2
+\frac{1}{3}V
\Big].
\label{ein_5d}
\eeq
Similarly, the scalar field equations read
\beq
&& \phi''+\frac{4a'}{a}\phi'
 =\phi\Big[2\lambda_3 \chi^2+\lambda_1\big(\phi^2-m_1^2\big)\Big],
\nonumber\\
&& \chi''+\frac{4a'}{a}\chi'
 =\chi\Big[2\lambda_3\phi^2+\lambda_2\big(\chi^2-m_2^2\big)\Big],
\eeq
We solve these equations under the boundary conditions shown
at the end of the previous section.
We will look for the thick de Sitter brane solution
with the asymptotically anti- de Sitter bulk.

\subsection{The six-dimensional model}

By substituting $n=2$ into our metric ansatz given by (\ref{metric_n}),
the general form of the six-dimensional metric is given by
\begin{eqnarray}
ds^2=a(r)^2 \gamma_{\mu\nu}dx^{\mu}dx^{\nu}-\lambda(r)(dr^2+r^2d\theta^2).
\end{eqnarray}
To find thick brane solutions, we solve the coupled Einstein-scalar system
given in Sec. III.
By substituting $n=2$ into the Einstein and scalar field equations
and combining them appropriately, we find the
equations
\beq
&& \frac{\lambda''}{\lambda}
-\Big(\frac{\lambda'}{\lambda}\Big)^2
-3\Big(\frac{a'}{a}\Big)^2
+\frac{3a'\lambda{}'}{a\lambda}
+\frac{1}{r}
\Big(
\frac{\lambda{}'}{\lambda}
+\frac{6a{}'}{a}
\Big)
+\frac{3\lambda}{a^2}
=-\frac{\epsilon}{2M_6^4}
\Big[\frac{1}{2}\Big(\phi{}'\Big)^2
    +\frac{1}{2}\Big(\chi{}'\Big)^2
    +\lambda V
\Big]\,,
\nonumber\\
&&-\frac{a{}''}{a}
  +\frac{a{}'}{ar}
  +\frac{a{}'\lambda{}'}{a\lambda}
=\frac{\epsilon}{2}
\Big[\frac{1}{2}(\phi{}'){}^2
   + \frac{1}{2}(\chi{}'){}^2
\Big],
\nonumber\\
&& \phi{}''+\Big(\frac{4a{}'}{a}+\frac{1}{r}\Big)\phi{}'
 =\lambda
   \phi\Big[2{\lambda}_3 \chi^2
+{\lambda}_1
    \big(\phi^2-m_1^2\big)\Big],
\nonumber\\
&& \chi{}''
+\Big(\frac{4a{}'}{a}+\frac{1}{r}\Big)\chi{}'
 =\lambda \chi
  \Big[2{\lambda}_3\phi^2
   +{\lambda_2}\big(\chi^2-m_2^2\big)\Big].
\label{6d}
\eeq
We solve these equations under the boundary conditions shown
at the end of the previous section.
We will look for the thick de Sitter brane solution
with the asymptotically flat bulk.
%

\subsection{The seven-dimensional model}

Similarly to the case of the six-dimensional model,
we consider the seven-dimensional model.
The bulk metric is given by Eq. (\ref{metric_n}) with $n=3$.
Then, by combining a set of equations appropriately,
the equations of motion can be obtained as follows
\beq
&&\frac{\lambda{}''}{\lambda}
-\frac{3}{5}\Big(\frac{\lambda'}{\lambda}\Big)^2
-\frac{12}{5}\Big(\frac{a{}'}{a}\Big)^2
+\frac{4}{5r}
\Big(\frac{6a{}'}{a}+\frac{13}{4}\frac{\lambda{}'}{\lambda}\Big)
+\frac{12a'\lambda{}'}{5a\lambda}
+\frac{12 \lambda}{5a^2}
=-\frac{\epsilon}{5}
 \Big[\big(\phi{}'\big)^2
     +\big(\chi{}'\big)^2
      +2V
 \Big]\,,
\nonumber\\
&&8\Big(\frac{a{}''}{a}
       -\frac{a{}'}{ar}
       -\frac{a{}'\lambda{}'}{a\lambda}
\Big)
+\frac{\lambda{}''}{\lambda}
-\frac{3}{2}\Big(\frac{\lambda{}'}{\lambda}\Big)^2
-\frac{\lambda{}'}{r \lambda}
=-2\epsilon
\Big[\big(\phi{}'\big)^2
     +\big(\chi{}'\big)^2\Big]\,.
\nonumber\\
&&\phi{}'{}'
+\Big(\frac{2}{r}
    +\frac{\lambda'}{2\lambda}
    +\frac{4 a'}{a}
\Big)
\phi'
=\lambda \phi
\Big[\lambda_1
\big(\phi^2-m_1^2\big)+2 \lambda_3 \chi^2\Big],
\nonumber\\
&&\chi{}''
+\Big(\frac{2}{r}
    +\frac{\lambda'}{2\lambda}
    +\frac{4 a'}{a}
\Big)
\chi{}'
=\lambda \chi
\Big[\lambda_2\big(\chi^2-m_2^2\big)
+2\lambda_3 \phi^2\Big]. \label{hagyou}
\eeq
The constraint equation is given by
\beq
&&
-\Big\{
-\frac{\lambda{}'{}^2}{4\lambda^2}
-\frac{\lambda{}'}{\lambda r}
-\frac{4 \lambda{}' a{}'}{\lambda a}
+\frac{6\lambda}{a^2}
-\frac{8a{}'}{ar}
-6\Big(\frac{a{}'}{a}\Big)^2
\Big\}
=\epsilon
\Big(
 \frac{1}{2}\phi{}'{}^2
+\frac{1}{2}\chi{}'{}^2
-\lambda V
\Big).
\eeq
As in the five and six-dimensional models,
we will solve them under the boundary conditions shown
at the end of the previous section.
We will look for the thick de Sitter brane solution
with asymptotically flat bulk.

\subsection{Numerics and solutions}

Let us present now the numerical examples of thick de Sitter brane solutions for the case
of phantom scalar field $\epsilon=-1$.
We solved the coupled Einstein-scalar system
by the iteration method. The detailed description of the method can be found, for example, in Ref.~\cite{2scalar2}.
The essence of this procedure is the following: on the first step one can solve the equation for the scalar field $\phi$
with some arbitrary selected function $\chi$ looking for a regular solution existing only at some value of the
parameter $m_1$. On this step the influence of gravitation is not taken into account. Then one inserts this solution for the
function $\phi$ into the equation for $\chi$ and
searches for a value of the parameter
$m_2$ yielding a regular solution. This procedure is repeated (three times is usually enough) for obtaining
acceptable convergence of values of the parameters $m_1, m_2$.
The obtained functions $\phi, \chi$ are inserting into the Einstein equations.
Equation \eqref{Einstein-nb}, which is the constraint equation, is using for specifying the boundary conditions
(see above). The obtained solutions for the metric functions $\lambda, a$ are inserted into the equations for
the scalar fields, and they are solved again for a search of eigenvalues of the
parameters $m_1, m_2$ with an account of gravitation. This procedure is repeated
as many times as it is necessary for
obtaining of acceptable convergence of values of the parameters $m_1, m_2$.

We set the parameters as
$\lambda_1=0.1$,
$\lambda_2=1.0$,
$\lambda_3=1.0$. For the five-dimensional case we have chosen
$\phi(0)=1.0$,
$\chi(0)=\sqrt{0.6}$
and
$\Lambda=-3.6$. For these parameters
we solved the nonlinear eigenvalue problem for $m_1$ and $m_2$.
After 11 iterations, the eigenvalues
$m_1\approx 1.989512$
and
$m_2\approx 1.964764$ were found.
In Fig. 1-4, we showed our numerical solutions.
The spacetime is asymptotically an anti-de Sitter one:  the asymptotic value
of the potential~\eqref{pot2} $\epsilon V_{\infty}=\epsilon(\lambda_2 m_2^4/4+ \Lambda)<0$
plays the role of a negative cosmological constant.

\begin{figure}
\begin{minipage}[t]{.45\textwidth}
   \begin{center}
    \includegraphics[scale=.75]{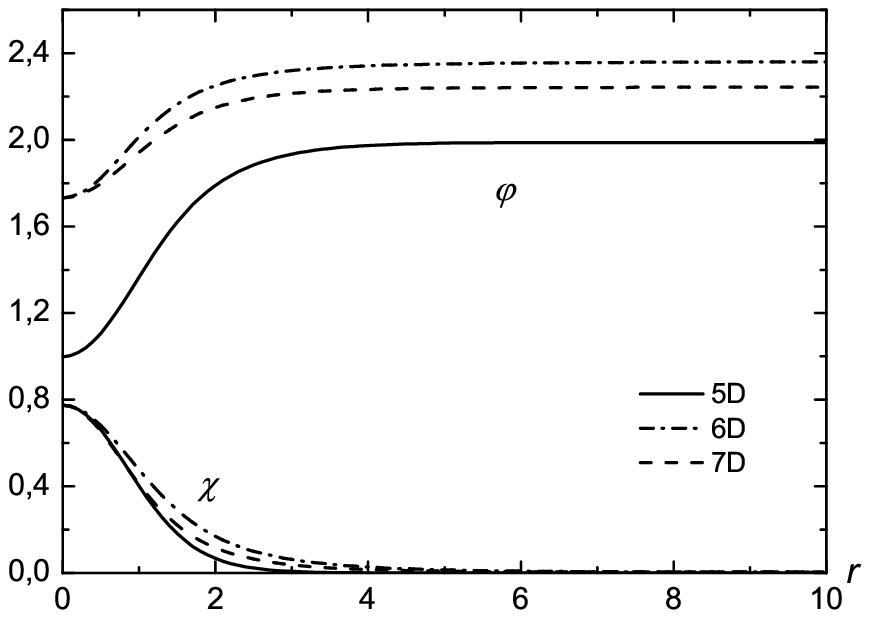}
\vspace{-.9cm}
        \caption{ The scalar field configurations $\phi$ and $\chi$
are shown
as functions of the dimensionless $r$.
}
   \end{center}
 \end{minipage}
\hspace{0.4cm}
\begin{minipage}[t]{.45\textwidth}
   \begin{center}
    \includegraphics[scale=.75]{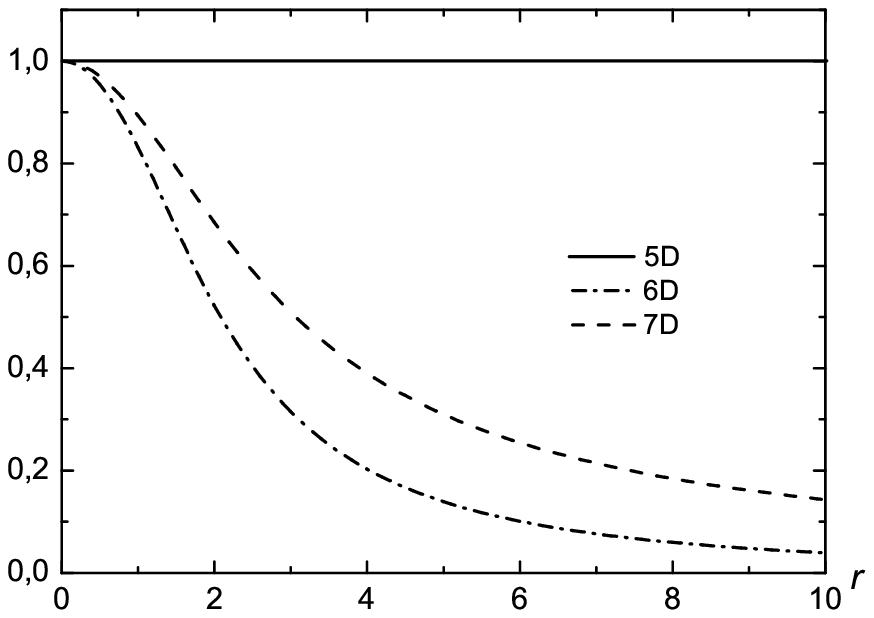}
\vspace{-.9cm}
        \caption{The metric function $\lambda(r)$ is shown
as functions of dimensionless $r$.
}
   \end{center}
   \end{minipage}
\end{figure}

\begin{figure}
\begin{minipage}[t]{.45\textwidth}
   \begin{center}
    \includegraphics[scale=.75]{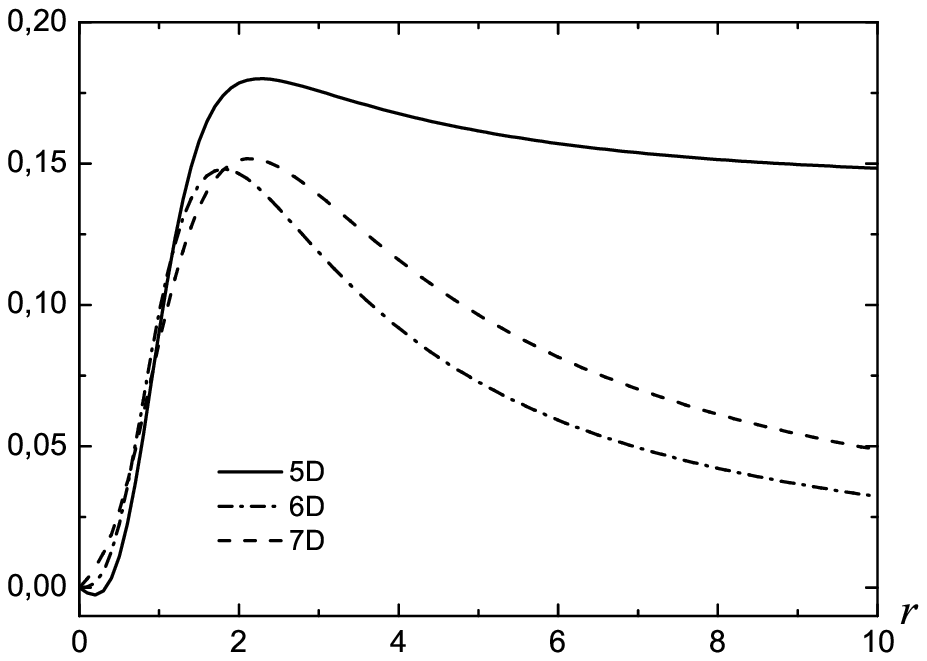}
\vspace{-.8cm}
        \caption{The function $a'/a$ is shown
as functions of dimensionless $r$.
}
   \end{center}
   \end{minipage}
\hspace{0.4cm}
\begin{minipage}[t]{.45\textwidth}
   \begin{center}
    \includegraphics[scale=.80]{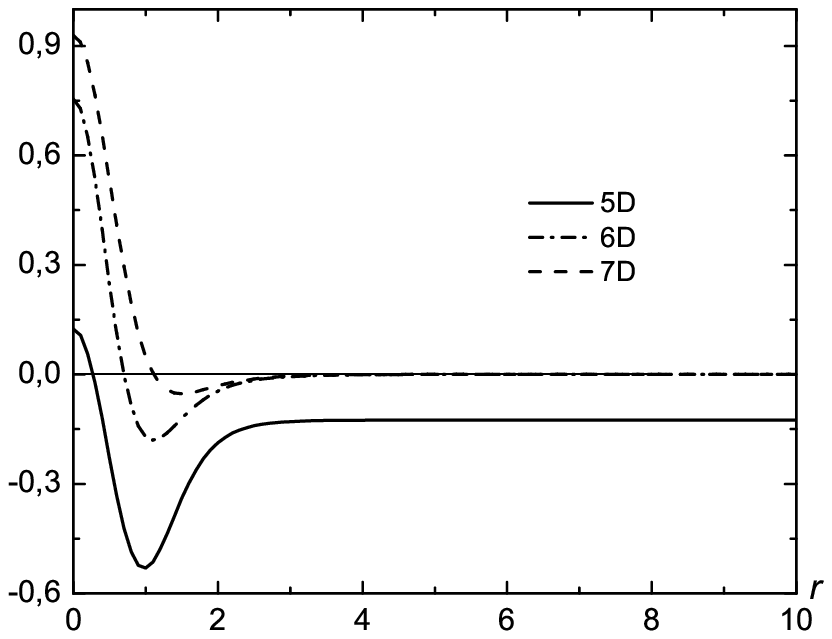}
\vspace{-.8cm}
        \caption{The energy density $T_0^0=\epsilon\left[\frac{1}{2\lambda}\left(\phi^{\prime 2}+\chi^{\prime 2}\right)+V\right]$ is shown
as functions of dimensionless $r$.
}
   \end{center}
   \end{minipage}
\end{figure}
For the six-dimensional case we take the boundary condition $\phi(0)=\sqrt{3}\approx 1.73205$
and
$\chi(0)=\sqrt{0.6}\approx 0.774597$.
We also choose the bulk cosmological constant as $\Lambda=-\lambda_2m_2^4/4$,
in order for our solutions to be asymptotically flat in the bulk.
After ten iterations, we find the eigenvalues
$m_1\approx 2.3590$ and
$m_2\approx 3.0599$.
In Fig. 1-4
we showed our numerical solution for a given set of parameters.

As for the case of the seven-dimensional case
we also choose the bulk cosmological constant as $\Lambda=-\lambda_2m_2^4/4$.
After ten iterations, we find the eigenvalues
$m_1\approx 2.24633$ and
$m_2\approx 3.11911$. The obtained solutions for this case are presented in Figs. 1-4.

\section{Stability}

For stability analysis,
let us write down the scalar field and Einstein equations,
by keeping the time-dependence of the metric and field functions.
Using the same rescaling as above and introducing new metric functions
$$
a=e^\alpha,\quad \lambda=e^\beta,
$$
one can get the following $(r,t)$ and $(r,r)$ components of
the Einstein equations
\begin{eqnarray}
\label{ein_time_1_n}
&&3\dot\alpha'-\frac{n+2}{2}\alpha'\dot\beta
+\frac{n-1}{2}\dot\beta'=
\epsilon\left[-\dot\phi \phi'-\dot\chi\chi'\right]e^{2\beta},\\
&&
(n-1)
\Big\{
e^{-2\alpha}
\Big(
   \dot{\alpha}\dot{\beta}
 +\frac{1}{2}\big(\ddot{\beta}+\dot{\beta}^2\big)
 +\frac{3}{2}H\dot{\beta}
\Big)
-2\big(\beta' +\frac{2}{r}\big)\alpha' e^{-\beta}
-\frac{n-2}{8}\big(\beta' +\frac{4}{r}\big)\beta' e^{-\beta}
\Big\}
-e^{-2\alpha}\dot{\beta}^2
\nonumber\\
&+&12e^{-2\alpha}
\big(\ddot{\alpha}
   +\dot{\alpha}^2
   +2
   -2\alpha{}'{}^2
     e^{2\alpha-\beta}
    +3\dot{\alpha}
\big)
=\epsilon
\Big[
-\frac{1}{2}e^{2\alpha}
   \Big(
   \dot{\phi}^2
  +\dot{\chi}^2
   \Big)
-\frac{1}{2}e^{\beta}
  \Big(
   \phi{}'{}^2
  +\chi{}'{}^2
   \Big)
+V
\Big].
\label{ein_time_2_n}
\end{eqnarray}
Here the dot refers to derivative with respect to dimensionless time $\tau=H t$.
The scalar filed equations are:
\begin{eqnarray}
\label{sfe_time_1}
&&e^{-2\alpha}
\ddot{\phi}
-e^{-\beta}
\phi''+
e^{-2\alpha}
\Big(
2\dot{\alpha}
+3
+\frac{n\dot{\beta}}{2}
\Big)\dot\phi
-e^{-\beta}
\Big(4\alpha'
+\frac{n-1}{r}
+\frac{n-2}{2}\beta'
\Big)
\phi'
=-\phi
\Big[
2\lambda_3\chi^2
+\lambda_1\big(\phi^2-m_1^2\big)
\Big]~,
\label{sfe_time_1}
\\
&&e^{-2\alpha}
\ddot{\chi}
-e^{-\beta}
\chi''+
e^{-2\alpha}
\Big(
2\dot{\alpha}
+3
+\frac{n\dot{\beta}}{2}
\Big)\dot\chi
-e^{-\beta}
\Big(4\alpha'
+\frac{n-1}{r}
+\frac{n-2}{2}\beta'
\Big)
\chi'
=-\chi
\Big[
2\lambda_3\phi^2
+\lambda_2\big(\chi^2-m_2^2\big)
\Big]~.
\label{sfe_time_2}
\end{eqnarray}
Now we search for regular perturbed solutions of the
above equations in the form
\begin{eqnarray}
\alpha&=&\alpha_0(r)+\delta\alpha(\tau,r),\quad
\beta=\beta_0(r)+\delta\beta(\tau,r),
\nonumber\\
\phi&=&\phi_0(r)+\delta\phi(\tau,r),\quad
\chi=\chi_0(r)+\delta\chi(\tau,r),
\end{eqnarray}
where the subscript 0 refers to static solutions of the equations
and
$\delta\alpha(\tau,r),
\delta\beta(\tau,r),
\delta\phi(\tau,r),
\delta\chi(\tau,r)$
are perturbations. Then we will have
from \eqref{ein_time_1_n} and \eqref{ein_time_2_n}:
\begin{eqnarray}
\label{eins_3}
&&3(\dot{\delta\alpha})'
-\frac{n+2}{2}\alpha_0^\prime\dot{\delta\beta}
+\frac{n-1}{2}(\dot{\delta\beta})'=
-\epsilon\left(\dot{\delta\phi}\phi_0^\prime+\dot{\delta\chi}\chi_0^\prime\right)e^{2\beta_0},\\
\label{eins_4}
&&(n-1)
\Big\{\frac{e^{-2\alpha_0}}{2}
\Big(
\ddot{\delta \beta}+3\dot{\delta\beta}
\Big)
-2e^{-\beta_0}
\Big[
-\alpha_0{}'\left(\beta_{0} {}'+\frac{2}{r}\right)\delta\beta
+\left(\beta_0{}'+\frac{2}{r}\right)\delta\alpha{}'
+\alpha_0{}'\delta\beta{}'
\Big]
\nonumber\\
&-&\frac{n-2}{8}e^{-\beta_0}
\Big[
-\beta_0{}'\left(\beta_0{}'+\frac{4}{r}\right)\delta\beta
+2\left(\beta_0{}'+\frac{2}{r}\right)\delta\beta{}'
\Big]
\Big\} \nonumber \\
&&+12e^{-2\alpha_0}\left[\left(\ddot{\delta\alpha}+3\dot{\delta\alpha}\right)-e^{-\beta_0}\left(2\alpha_0^\prime\delta\alpha'-
\alpha_0^{\prime 2}\delta\beta
-2\alpha_0^{\prime 2}\delta\alpha\right)
-4\delta\alpha\right] \nonumber \\
&&=\epsilon
\Big[
-e^{\beta_0}
\big(\phi_0{}'\delta\phi{}'
  +\chi_0{}'\delta\chi{}'
\big)
-
\frac{e^{\beta_0} \delta\beta}{2}
\big(\phi_0{}'{}^2
  +\chi_0{}'{}^2
\big)
\nonumber\\
&+&
 \lambda_1\phi_0\big(\phi_0^2-m_1^2\big)\delta \phi
+\lambda_2\chi_0\big(\chi_0^2-m_2^2\big)\delta \chi
+2\lambda_3 \phi_0^2\chi_0\delta\chi
+2\lambda_3 \chi_0^2\phi_0\delta\phi
\Big],
\end{eqnarray}
and from \eqref{sfe_time_1} and \eqref{sfe_time_2} follows
\begin{eqnarray}
\label{sca_1}
&&e^{-2\alpha_0}
\big(\ddot{\delta\phi}+3\dot{\delta\phi}\big)
-e^{-\beta_0}
\big(
\delta \phi{}''
-\delta\beta\phi_0{}''
\big)
-e^{-\beta_0}
\Big[
\Big(4\alpha_0{}'+\frac{n-2}{2}\beta_0{}'+\frac{n-1}{r}\Big)
 \big(\delta\phi{}'-\delta\beta \phi_0{}'\big)
\nonumber\\
&&+\phi_0{}'
\Big(4\delta\alpha{}'+\frac{n-2}{2}\delta\beta{}'\Big)
\Big]
=-4\lambda_3\phi_0\chi_0\delta\chi
-\Big[2\lambda_3\chi_0^2+\lambda_1
\big(3\phi_0^2-m_1^2\big)\Big]\delta\phi,\\
&&\label{sca_2}
e^{-2\alpha_0}
\big(\ddot{\delta\chi}+3\dot{\delta\chi}\big)
-e^{-\beta_0}
\big(
\delta \chi{}''
-\delta\beta\chi_0{}''
\big)
-e^{-\beta_0}
\Big[
\Big(4\alpha_0{}'+\frac{n-2}{2}\beta_0{}'+\frac{n-1}{r}\Big)
 \big(\delta\chi{}'-\delta\beta \chi_0{}'\big)
\nonumber\\
&+&\chi_0{}'
\Big(4\delta\alpha{}'+\frac{n-2}{2}\delta\beta{}'\Big)
\Big]
=-4\lambda_3\phi_0\chi_0\delta\phi-\left[2\lambda_3\phi_0^2+\lambda_2
\left(3\chi_0^2-m_2^2\right)\right]\delta\chi~.
\end{eqnarray}
One can see that Eqs.~\eqref{eins_4}-\eqref{sca_2} contain similar terms like
$
\left(\ddot{\delta y}+3\dot{\delta y}\right),
$
where $\delta y$ is one of functions
$\delta\alpha(\tau,r),
\delta\beta(\tau,r),
\delta\varphi(\tau,r),
\delta\chi(\tau,r)$.
Such term implies that one can search for a solution in the form:
$
{\delta y=y_1(r)e^{-\frac{3}{2}\tau}\cos{\omega \tau},}
$
where $\omega=\sqrt{\omega_0^2-\frac{9}{4}}>0$. It implies that $\omega_0>\frac{3}{2}$. Inserting the last expression into
\eqref{eins_3}-\eqref{sca_2}, one can get
\begin{eqnarray}
&&\label{eins_5}
3\alpha_1^\prime
-\frac{n+2}{2}\alpha_0^\prime\beta_1
+\frac{n-1}{2}\beta_1^\prime=
-\epsilon
\big(\phi_0^\prime\phi_1
+\chi_0^\prime\chi_1
\big)e^{2\beta_0},\\
\label{eins_6}
&&(n-1)
\Big\{-\frac{e^{-2\alpha_0}}{2}\omega_0^2 \beta_1
-2e^{-\beta_0}
\Big[
-\alpha_0{}'\left(\beta_{0} {}'+\frac{2}{r}\right)\beta_1
+\left(\beta_0{}'+\frac{2}{r}\right)\alpha_1 {}'
+\alpha_0{}'\beta_1{}'
\Big]
\nonumber\\
&-&\frac{n-2}{8}e^{-\beta_0}
\Big[
-\beta_0{}'\left(\beta_0{}'+\frac{4}{r}\right)\beta_1
+2\left(\beta_0{}'+\frac{2}{r}\right)\beta_1{}'
\Big]
\Big\}
+12 e^{-2\alpha_0}
\Big[
-\omega_0^2\alpha_1
\nonumber\\
&-& e^{-\beta_0}\left(2\alpha_0^\prime\alpha_1^\prime-
\alpha_0^{\prime 2}\beta_1-2\alpha_0^{\prime 2}\alpha_1\right)-4 \alpha_1
\Big]
=\epsilon
\Big[
-e^{\beta_0}
\big(\phi_0{}'\phi_1{}'
  +\chi_0{}'\chi_1{}'
\big)
-
\frac{e^{\beta_0} \beta_1}{2}
\big(\phi_0{}'{}^2
  +\chi_0{}'{}^2
\big)
\nonumber\\
&+& \lambda_1\phi_0\big(\phi_0^2-m_1^2\big)\phi_1
+\lambda_2\chi_0\big(\chi_0^2-m_2^2\big)\chi_1
+2\lambda_3 \phi_0^2\chi_0\chi_1
+2\lambda_3 \chi_0^2\phi_0\phi_1
\Big],
\end{eqnarray}
and
\begin{eqnarray}
\label{sca_3}
&&-e^{-2\alpha_0}\omega_0^2\phi_1
-e^{-\beta_0}
\big(\phi_1{}''
-\beta_1\phi_0{}''
\big)
-e^{-\beta_0}
\Big[
\Big(4\alpha_0{}'+\frac{n-2}{2}\beta_0{}'+\frac{n-1}{r}\Big)
 \big(\phi_{1}{}'-\beta_1 \phi_0{}'\big)
\nonumber\\
&+&\phi_0{}'
\Big(4\alpha_1{}'+\frac{n-2}{2}\beta_1{}'\Big)
\Big]
=-4\lambda_3\phi_0\chi_0\chi_1-\Big[2\lambda_3\chi_0^2+\lambda_1
\big(3\phi_0^2-m_1^2\big)\Big]\phi_1,
\\
&&\label{sca_4}
-\omega_0^2 e^{-2\alpha_0}\chi_1
-e^{-\beta_0}
\big(
\chi_1{}''
-\beta_1\chi_0{}''
\big)
-e^{-\beta_0}
\Big[
\Big(4\alpha_0{}'+\frac{n-2}{2}\beta_0{}'+\frac{n-1}{r}\Big)
 \big(\chi_1{}'-\beta_1 \chi_0{}'\big)
\nonumber\\
&+&\chi_0{}'
\Big(4\alpha_1{}'+\frac{n-2}{2}\beta_1{}'\Big)
\Big]
=-4\lambda_3\phi_0\chi_0\phi_1
-\Big[2\lambda_3\phi_0^2+\lambda_2
\big(3\chi_0^2-m_2^2\big)\Big]\chi_1.
\end{eqnarray}
So we have four usual differential equations for perturbations.
For existence of stable solutions we need to provide positiveness of an eigenvalue $\omega_0^2$.
To do this we  will search
for finite solutions of above equations.

For the five-dimensional model, by setting $\beta_1=0$ (and $\beta_0=0$)
and substituting $n=1$,
the perturbations equations are simply reduced to
\beq
&&\phi_1{}''
+4\alpha_0{}'\phi_1{}'
-\Big[
2\lambda_3\chi_0^2
+\frac{4\epsilon}{3}\phi_0{}'{}^2
+\lambda_1
\left(3\phi_0^2-m_1^2\right)
\Big]
\phi_1
-\Big(
\frac{4\epsilon}{3}\phi_0{}'\chi_0{}'
+4\lambda_3\phi_0\chi_0
\Big)\chi_1
+\omega_0^2 e^{-2\alpha_0} \phi_1
=0,\label{p51}
\nonumber
\\
&&
\chi_1{}''
+4\alpha_0{}'\chi_1{}'
-\Big[
2\lambda_3\phi_0^2
+\frac{4\epsilon}{3}\chi_0{}'{}^2
+\lambda_2
\left(3\chi_0^2-m_2^2\right)
\Big]
\chi_1
-\Big(
\frac{4\epsilon}{3}\phi_0{}'\chi_0{}'
+4\lambda_3 \phi_0\chi_0
\Big)\phi_1
+\omega_0^2 e^{-2\alpha_0} \chi_1
=0,
\label{p52}
\eeq
where we have used Eq.~\eqref{eins_5}
$$
3\alpha_1^\prime=
-\epsilon\left(\phi_0^\prime\phi_1+\chi_0^\prime\chi_1\right)
$$
to eliminate $\alpha_1{}'$.
We  solved numerically the Eqs.~(\ref{p51}), using
the background  solutions obtained in Sec. III, with the boundary conditions
\beq
\phi_1(0)=1.0,
\quad
\phi_1^\prime(0)=0.0,
\quad
\chi_1(0)=-1.0,
\quad
\chi_1^\prime(0)=0.0,
\quad
\phi_1(r\to\infty)\to 0,
\quad
\chi_1(r\to \infty)\to 0.
\eeq
In the case of the thick Minkowski brane solutions supported by the
phantom scalar fields, it has been shown that these solutions are stable
\cite{phantom}.
In our case,
the perturbation
equations also have regular solutions.
Also, we have confirmed that there are no solutions
with $\omega_0^2<0$, i.e.  unstable solutions are absent.
In Fig 5, we showed an example of numerical solutions for
perturbations with $\omega_0\approx 13.179575$.
We conclude that our thick de Sitter brane solutions supported by the
phantom scalar fields are stable,
irrespectively of the value of the Hubble parameter $H$.
As a special limit $H\to 0$, we may recover the result
for the Minkowski brane solutions.
\begin{figure}
\begin{minipage}[t]{.50\textwidth}
   \begin{center}
    \includegraphics[scale=.85]{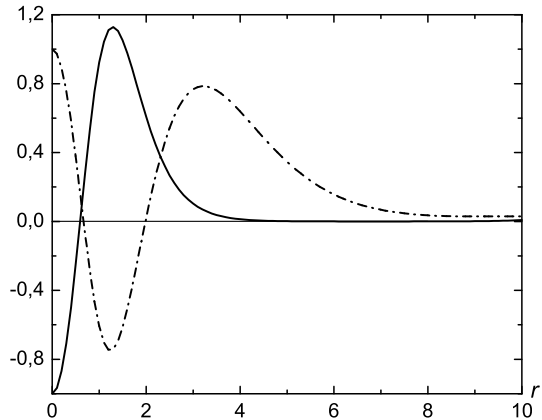}
\vspace{-.8cm}
        \caption{ The perturbed scalar field configurations
$\phi_1$ (solid curve) and $\chi_1$
for $\omega_0\approx 13.179575$
(dashed-dotted curve)
are shown
as functions of $r$.
As the background solutions,
we used the one shown in Figs. 1-3.}
   \end{center}
   \end{minipage}
\end{figure}

It is also reasonable to expect that
our phantom thick brane solutions in six- and seven-dimensional
spacetime are also completely stable,
since the spacetime structures of higher-dimensional solutions
are essentially very similar to that of the five-dimensional
solution.

\section{Summary and Discussions}

In this paper, we have presented five-,
six-, and seven-dimensional thick de Sitter brane solutions
supported by two interacting ({\it phantom}) scalar fields.
The special form of the potential~\eqref{pot2} allows us to find regular solutions
with finite energy density. It was shown that asymptotically there exist anti-de Sitter spacetime
for the five-dimensional case and flat spacetime for the six- and seven-dimensional cases.

Also
we have investigated the stability of our thick brane solutions.
In the five-dimensional thick de Sitter brane solutions,
we have shown the presence of decaying solutions for perturbations.
We also could not find any mode with negative eigenvalue.
This, of course, implies that our solutions in five-dimensional
spacetime are stable.
It is also quite reasonable to expect
that our thick brane solutions in higher dimensions are
also stable since the spacetime and vacuum structures are essentially the same
as in the case of five spacetime dimensions.


\section*{Acknowledgements}

The work of M. M. was supported
by the project TRR 33 {\it The Dark Universe} at the ASC.
VD is grateful to the Alexander von Humboldt Foundation for financial
support and to V. Mukhanov for invitation to Universitaet Muenchen for
research.
This work was supported by the Korea Science and Engineering Foundation(KOSEF)
grant funded by the Korea government(MEST) through the Center for Quantum Spacetime(CQUeST)
of Sogang University with grant number R11 - 2005 - 021.




\begin{thebibliography}{999}



\bibitem{rs}
  L.~Randall and R.~Sundrum,
  Phys.\ Rev.\ Lett.\  {\bf 83}, 3370 (1999)
  [arXiv:hep-ph/9905221];
 L.~Randall and R.~Sundrum,
  Phys.\ Rev.\ Lett.\  {\bf 83}, 4690 (1999)
  [arXiv:hep-th/9906064].



\bibitem{bw}
  R.~Maartens,
  Living Rev.\ Rel.\  {\bf 7}, 7 (2004)
  [arXiv:gr-qc/0312059]l
  P.~Brax, C.~van de Bruck and A.~C.~Davis,
  Rept.\ Prog.\ Phys.\  {\bf 67}, 2183 (2004)
  [arXiv:hep-th/0404011].




%



\bibitem{tb_gen}
  O.~DeWolfe, D.~Z.~Freedman, S.~S.~Gubser and A.~Karch,
  Phys.\ Rev.\ D {\bf 62}, 046008 (2000)
  [arXiv:hep-th/9909134];
  M.~Gremm,
  Phys.\ Lett.\ B {\bf 478}, 434 (2000)
  [arXiv:hep-th/9912060];
  C.~Csaki, J.~Erlich, T.~J.~Hollowood and Y.~Shirman,
  Nucl.\ Phys.\ B {\bf 581}, 309 (2000)
  [arXiv:hep-th/0001033];
  M.~Giovannini,
  Phys.\ Rev.\ D {\bf 65}, 064008 (2002)
  [arXiv:hep-th/0106131].





\bibitem{thick}
  S.~Kobayashi, K.~Koyama and J.~Soda,
  Phys.\ Rev.\ D {\bf 65}, 064014 (2002)
  [arXiv:hep-th/0107025].
  N.~Sasakura,
  JHEP {\bf 0202}, 026 (2002)
  [arXiv:hep-th/0201130];
  N.~Sasakura,
  Phys.\ Rev.\ D {\bf 66}, 065006 (2002)
  [arXiv:hep-th/0203032];
  A.~Z.~Wang,
  Phys.\ Rev.\ D {\bf 66}, 024024 (2002)
  [arXiv:hep-th/0201051];
  K.~A.~Bronnikov and B.~E.~Meierovich,
  Grav.\ Cosmol.\  {\bf 9}, 313 (2003)
  [arXiv:gr-qc/0402030];
  M.~Minamitsuji, W.~Naylor and M.~Sasaki,
  Nucl.\ Phys.\  B {\bf 737}, 121 (2006)
  [arXiv:hep-th/0508093];
M.~Minamitsuji, W.~Naylor and M.~Sasaki,
  Phys.\ Lett.\  B {\bf 633}, 607 (2006)
  [arXiv:hep-th/0510117];
 N.~Barbosa-Cendejas and A.~Herrera-Aguilar,
  Phys.\ Rev.\  D {\bf 73}, 084022 (2006)
  [Erratum-ibid.\  D {\bf 77}, 049901 (2008)]
  [arXiv:hep-th/0603184].

\bibitem{2scalar1}
  V.~Dzhunushaliev,
  Grav.\ Cosmol.\  {\bf 13}, 302 (2007)
  [arXiv:gr-qc/0603020];


\bibitem{phantom}
 V.~Dzhunushaliev, V.~Folomeev, S.~Myrzakul and R.~Myrzakulov,
  arXiv:0804.0151 [gr-qc].




\bibitem{2scalar2}
 V.~Dzhunushaliev, V.~Folomeev, K.~Myrzakulov and R.~Myrzakulov,
  arXiv:0705.4014 [gr-qc];
 V.~Dzhunushaliev, V.~Folomeev, D.~Singleton and S.~Aguilar-Rudametkin,
  Phys.\ Rev.\  D {\bf 77}, 044006 (2008)
  [arXiv:hep-th/0703043].



\bibitem{higher_dim}
  M.~Giovannini, H.~Meyer and M.~E.~Shaposhnikov,
  Nucl.\ Phys.\  B {\bf 619}, 615 (2001)
  [arXiv:hep-th/0104118];
T.~Gherghetta, E.~Roessl and M.~E.~Shaposhnikov,
  Phys.\ Lett.\  B {\bf 491}, 353 (2000)
  [arXiv:hep-th/0006251].
R.~S.~Torrealba S.,
arXiv:0803.0313 [hep-th].







\bibitem{gauge}
  V.~Dzhunushaliev,
  arXiv:hep-ph/0605070.
\bibitem{heisenberg}
W. Heisenberg,
\textit{Introduction to the unified field theory of elementary particles.},
(Max - Planck - Institut f\"ur Physik und Astrophysik, Interscience
Publishers London, New York-Sydney, 1966).


\bibitem{Singleton}
D. Singleton, {\it Phys.Rev.} {\bf D70}, 065013 (2004).

\end{thebibliography}
\end{document}